\documentclass[aps, pre, twocolumn, 10pt, amsmath, amssymb, amsfonts]{revtex4-1}

\usepackage{graphicx}
\usepackage[usenames]{color}
\usepackage[colorlinks=True,
            urlcolor=blue,
            linkcolor=blue,
            citecolor=blue]{hyperref}

\newlength{\figwidth}
\newlength{\figheight}
\newcommand{\ud}{\,\mathrm{d}}
\newcommand{\vect}[1]{\mathbf{#1}}

\begin{document}
\title{Scaling of the dynamics of flexible Lennard-Jones chains.\\II. Effects of harmonic bonds}
\author{Arno A. Veldhorst}
\email{a.a.veldhorst@gmail.com}
\altaffiliation{Current Address: Laborat\'orio de Espectroscopia Molecular, Instituto de Qu\'imica, Universidade de S\~ao Paulo, CP 26077, CEP 05513-970 S\~ao Paulo, SP, Brazil}
\affiliation{DNRF Center ``Glass and Time'', IMFUFA, Dept. of Sciences, Roskilde University, P.O. Box 260, DK-4000 Roskilde, Denmark}
\author{Jeppe C. Dyre}
\affiliation{DNRF Center ``Glass and Time'', IMFUFA, Dept. of Sciences, Roskilde University, P.O. Box 260, DK-4000 Roskilde, Denmark}
\author{Thomas B. Schr{\o}der}
\email{tbs@ruc.dk}
\affiliation{DNRF Center ``Glass and Time'', IMFUFA, Dept. of Sciences, Roskilde University, P.O. Box 260, DK-4000 Roskilde, Denmark}
\date{\today}

\begin{abstract}
The previous paper [Veldhorst \textit{et al.}, J. Chem. Phys. \textbf{141}, 054904 (2014)] demonstrated that the isomorph theory explains the scaling properties of a liquid of flexible chains consisting of ten Lennard-Jones particles connected by rigid bonds. We here investigate the same model with harmonic bonds. The introduction of harmonic bonds almost completely destroys the correlations in the equilibrium fluctuations of the potential energy and the virial. According to the isomorph theory, if these correlations are strong a system has isomorphs, curves in the phase diagram along which structure, dynamics and the excess entropy are invariant. The Lennard-Jones chain liquid with harmonic bonds \emph{does} have curves in the phase diagram along which the structure and dynamics are invariant. The excess entropy is \emph{not} invariant on these curves, which we refer to as ``pseudoisomorphs''. In particular this means that Rosenfeld's excess-entropy scaling (the dynamics being a function of excess entropy only) does not apply for the Lennard-Jones chain with harmonic bonds.
\end{abstract}

 \maketitle 
\section{Introduction}
The dynamics of a viscous liquid near the glass transition is very state-point dependent. Relatively small changes in temperature or pressure can change the relaxation time and the viscosity significantly. The dynamics of a liquid are usually dependent on two state variables, and this state-point dependence is material specific. A general understanding of the state-point dependence of the dynamics of liquids has since a long time been a goal in field of  liquid-state and glass physics~\cite{Debenedetti2001, Dyre2006, Ediger2012}.

Experimental results by T{\"o}lle \textit{et al.}~\cite{Tolle1998, Tolle2001} indicated that the problem of what controls the relaxation time of viscous liquids might be simplified, if instead of pressure $p$ and temperature $T$, one focuses on density $\rho$  and temperature. It was found that the dynamics of \textit{ortho}-terphenyl, measured at different state points can be collapsed onto a single curve by plotting it as a function of $h(\rho)/T$, with $h(\rho)=\rho^4$. Later results have shown that the dynamics of many liquids can be collapsed when plotted as a function of $h(\rho)/T$, albeit with a material-dependent $h(\rho)$~\cite{Alba-Simionesco2002, Dreyfus2003, Casalini2004}. A review by Roland \textit{et al.}~\cite{Roland2005} established that for many liquids $h(\rho)$ is well approximated by a power law of the density, $\rho^{\gamma_s}$, with $\gamma_s$ being a material specific scaling parameter. We refer to this as \emph{power-law} density scaling. The fact that the dynamics of some liquids were found to be a function of the combined variable $\rho^{\gamma_s}/T$ indicated that there might be a single underlying quantity that ``controls'' the dynamics. 

Some liquids  have strong correlations in the equilibrium fluctuations of the energy and pressure~\cite{Pedersen2008, paper1, paper2}. Specifically, these correlations are found in the configurational parts of the energy $E$ and pressure $p$, i.e., the potential energy $U$ and the virial $W$, respectively. These only depend on the positions of the particles $\vect{r}_i$, in contrast to the kinetic energy $K$ and the temperature, which only depend on the momenta of the particles $\vect{p}_i$:
\begin{align}
  E  &= K(\vect{p}_1, \ldots, \vect{p}_N) + U(\vect{r}_1, \ldots, \vect{r}_N)\,,\\
  pV &= N k_B T(\vect{p}_1, \ldots, \vect{p}_N) + W(\vect{r}_1, \ldots, \vect{r}_N)\,.
\end{align}
The $UW$ correlations are quantified by the standard Pearson correlation coefficient
\begin{equation}\label{eq:R}
  R = \frac{\left< \Delta W \Delta U \right>}
           {\sqrt{\left<(\Delta W)^2\right> \left<(\Delta U)^2\right>}}\,,
\end{equation}
where $\Delta$ denotes the difference from the mean ($\Delta U = U - \langle U \rangle$) and angular brackets denote the NVT ensemble average (constant number of particles, volume, and temperature). Liquids with $R\geqslant 0.9$ were initially called ``strongly correlating'', but since this sometimes led to confusion with strongly correlated quantum systems, they are now referred to as  ``Roskilde simple'' liquids or just ``R liquids''~\cite{Malins2013, Abramson2014, Flenner2014, Henao2014, Pieprzyk2014, Prasad2014, Buchenau2015, Heyes2015, Schmelzer2015}.

The discovery of this class of liquids subsequently led to the development of the isomorph theory~\cite{paper3, paper4}. The isomorph theory explains why R liquids have many properties that make them simpler than other liquids. The main prediction is that liquids that belong to this class have curves in their phase diagram called \emph{isomorphs} along which many properties are invariant. The invariance includes the dynamics, structure, and excess entropy (the entropy minus the entropy of the ideal gas at same temperature and density)~\cite{paper4}. These liquids thus have a phase diagram that is effectively one dimensional for many properties (but not for, e.g., the pressure and the free energy).

The isomorph theory provides a theoretical explanation for the empirical power-law density scaling. The theory does not assume anything about the functional form of $h(\rho)$, and it was indeed discovered that for many model liquids~\cite{Veldhorst2012, Ingebrigtsen2012b, Bailey2013, Bohling2014} and two real liquids~\cite{Bohling2012} that $h(\rho)$ is not well approximated by a power law if the change in density is larger than 10-20\%. Power-law density scaling is a good approximation when density changes are small, which is often the case in experiments. Simple power-law density scaling also works in the case of low-density supercritical fluids~\cite{Galliero2011, Delage-Santacreu2015}. Another advantage of the isomorph theory is that it provides a prediction for the (density-dependent) value of $\gamma_s$, which was previously treated as an empirical parameter. Thus, $\gamma_s$ can be estimated independently of the scaling procedure, although this is much more easily done in computer simulations~\cite{paper4, paper5, Bohling2014} than in experiments~\cite{Gundermann2011, Casalini2014}.

The isomorph theory is consistent with another scaling method, which was proposed by Rosenfeld~\cite{Rosenfeld1977} and later in a slightly different form by Dzugutov~\cite{Dzugutov1996}. In this scaling the dynamics of many liquids were found to be a function of the excess entropy. Since excess entropy and the dynamics are both invariant along isomorphs, this is in agreement with the isomorph theory, although the isomorph theory does not predict which function expresses the relaxation time in terms of the excess entropy.

Initially, the isomorph theory was tested on simple atomic model systems~\cite{paper4, Veldhorst2012} and small rigid molecules~\cite{Ingebrigtsen2012b}. However, liquids that have been shown to obey power-law density scaling are usually organic liquids with often internal degrees of freedom. In particular, many polymeric liquids obey power-law density scaling. This led us to investigate the applicability of the isomorph theory to flexible chain molecules in our previous publication (paper I)~\cite{Veldhorst2014}. We showed that the isomorph theory describes the properties of the flexible Lennard-Jones chains very well, except for small deviations due to intramolecular effects. These effects were due to the covalent bonds in the chain, which do not scale with density.

In paper~I~\cite{Veldhorst2014}, the covalent bonds were simulated using a constraint algorithm to keep the bond length fixed. This is not the most common way to simulate molecules with Molecular Dynamics~\cite{Leach}. In the field of (bio)chemistry, for instance, coarse-grained and all-atom models generally have force fields that model covalent bonds as harmonic springs~\cite{Cornell1995, Marrink2007, Brooks2009}. In this paper we show that the bond type has large implications on the applicability of the isomorph theory. We show that the Lennard-Jones chains with harmonic bonds have curves in their phase diagram along which the dynamics and structure are invariant. Using a method that is not dependent on the validity of the isomorph theory, we also generate curves of constant excess entropy (configurational adiabats). We find that unlike isomorphs, curves of invariant dynamics do not coincide with the configurational adiabats. We propose the name ``\emph{pseudo}isomorphs'' for curves that have invariant dynamics and structure, but not constant excess entropy. Our result has important consequences for the applicability of Rosenfeld's excess-entropy scaling We find that this scaling does not work for Lennard-Jones chain liquids with harmonic bonds.

The paper is structured as follows. In the next section we briefly review relevant aspects of the isomorph theory. We then introduce the simulation method and the Lennard-Jones chain model in Sec.~\ref{sec:model}, where we also show how the different bond types affect the dynamics and the $UW$ correlations of the liquid. We construct a configurational adiabat in Sec.~\ref{sec:configurational_adiabat} and curves of invariant dynamics in Sec.~\ref{sec:pseudoisomorph}, and test to which degree these curves resemble isomorphs. The findings are summarized in Sec.~\ref{sec:conclusion}.

\section{The isomorph theory\label{sec:theory}}
In the isomorph theory~\cite{paper4} uses reduced units, making quantities dimensionless using macroscopic quantities such as temperature and pressure. Quantities in reduced units are denoted by a tilde. Examples are the reduced distance $\tilde{r} = \rho^{1/3}r$, reduced energy $\tilde{U}=U/(k_BT)$, and reduced time $\tilde{t} = \rho^{1/3}(k_BT/m_a)^{1/2}t$, with $m_a$ being the average particle mass. Denoting a configuration as $\vect{R} = \left(\vect{r}_1, \ldots, \vect{r}_N \right)$, it is expressed in reduced units as $\tilde{\vect{R}} = \rho^{1/3}\vect{R}$ (which is equivalent to scaling the configuration to unit density).

At two state points with densities $\rho_1$ and $\rho_2$ pairs of configurations exist that have the same coordinates in reduced units
\begin{equation}\label{eq:reduced_conf}
  \rho_1^{1/3}\tilde{\vect{R}}_1 =  \rho_2^{1/3}\vect{R}_2 \equiv \tilde{\vect{R}}\,,
\end{equation}
i.e., they are scaled versions of each other. If the two state points have temperature $T_1$ and $T_2$, they are defined to be isomorphic if the Boltzmann factors of all pairs of scaled configurations obey~\cite{paper4}
\begin{equation}\label{eq:isom_def}
  \exp\left(-\dfrac{U(\vect{R}_1)}{k_BT_1}\right) = C_{12}\exp\left(-\dfrac{U(\vect{R}_2)}{k_BT_2}\right)\,,
\end{equation}
with the same constant $C_{12}$. In practice, this should hold to a good approximation for all physically relevant configurations.

From this definition it follows that the structure of the liquid is invariant in reduced units at isomorphic state points, because the relative probabilities of configurations that are scaled versions of each other are the same at both state points. Another important consequence of Eq.~(\ref{eq:isom_def}) is that the excess entropy $S_{ex}$ is the same at isomorphic  state points~\cite{paper4}. 

Taking the logarithm of Eq.~(\ref{eq:isom_def}) and expressing it in reduced units, one finds that
\begin{equation}\label{eq_isomorphDef-log-reduced}
  \tilde{U}(\vect{R}_1) = \tilde{U}(\vect{R}_2) - \ln(C_{1,2})\,.
\end{equation}
In other words, the potential energy landscapes at the two state points are simply scaled versions of each other. Defining the reduced force as the gradient of the reduced potential energy surface $\tilde{\vect{F}}=-\tilde{\nabla}\tilde{\vect{U}}$ with $\tilde{\nabla} = \rho^{-1/3}\nabla$, one finds that the forces at two isomorphic state points are the same in reduced units, $\tilde{\vect{F}}_1=\tilde{\vect{F}}_2$, for configurations obeying Eq.~(\ref{eq:reduced_conf}). A particle's reduced mass is given by $\tilde{m}_i = m_i/m_a$ where $m_a$ is the average particle mass. Using this, Newton's second law in reduced units, $\tilde{m}_i\ddot{\tilde{\vect{r}}}_i = \tilde{\vect{F}}_i$, leads to invariant dynamics when expressed in reduced units~\cite{paper4}.

Isomorphs can be generated using the property of constant excess entropy. This is done using the scaling exponent
\begin{equation}\label{eq:gamma-fluctuations}
  \gamma = \frac {\left<\Delta W \Delta U\right>} {\left<(\Delta U)^2\right>}\,.
\end{equation}
Using the standard fluctuation formulae it can be shown that $\left<\Delta W \Delta U\right>/\left<(\Delta U)^2\right> = (\partial\left<W\right>/\partial T)_V/(\partial \left<U\right>/\partial T)_V$~\cite{paper1}. This can then be shown to be equal to the slope of a configurational adiabat in the $(\log T, \log\rho)$ using the configurational version of the $V-T$ Maxwell relation~\cite{paper4}:
\begin{equation}\label{eq:gamma-rhoT}
  \gamma =\left(\frac{\partial \ln T}{\partial \ln \rho}\right)_{S_{ex}}\,.
\end{equation}
Equations (\ref{eq:gamma-fluctuations}) and (\ref{eq:gamma-rhoT}) can be used to map out a configurational adiabat for any system, using he fluctuations in $U$ and $W$. In practice this is done by calculating $\gamma$ from the fluctuations using Eq.~(\ref{eq:gamma-fluctuations}) and than calculating the temperature at another state point with a slightly different density using Eq.~(\ref{eq:gamma-rhoT}).

For a more in depth description of the isomorph theory the reader is referred to a recent feature article~\cite{Dyre2014}. It should also be noted that the isomorph theory was recently generalized by defining a Roskilde-simple system by the property that the order of potential energies is maintained for uniform scaling of configurations: $U(\vect{R}_a)<U(\vect{R}_b)\Rightarrow U(\lambda\vect{R}_a)<U(\lambda\vect{R}_b)$~\cite{Schroder2014}. For the properties considered in this paper the new formulation of the theory leads to the same predictions as the original formulation of the theory.

\section{Model and simulation method\label{sec:model}}
\subsection{Simulation method}
We simulated the flexible Lennard-Jones chain (LJC) model in the liquid phase. The chains consist of ten particles. All but the bonded particle pairs in the chains interact through the well-known Lennard-Jones potential
\begin{equation}
  \upsilon(r) 
    = 4\varepsilon\left[\left(\dfrac{r}{\sigma}\right)^{-12} 
                      - \left(\dfrac{r}{\sigma}\right)^{-6}\right]\,,
\end{equation}
cut and shifted at $2.5\sigma$. The interaction between bonded particle pairs is modeled by a harmonic spring
\begin{equation}
  \upsilon(r) = 0.5k(r-\sigma)^2
\end{equation}
with spring constant $k=3000\varepsilon/\sigma^2$. Note the bond length is the same as the distance at which the Lennard-Jones energy is zero. The simulation time step was 0.0025, which is sufficiently small for the large spring constant used. The simulations were carried out in the NVT ensemble, keeping the temperature fixed using a Nos{\'e}-Hoover thermostat. The Nos{\'e}-Hoover thermostat is known to not sample an harmonic potential correctly, but this does not affect the results for dense systems like the liquids studied here~\cite{Toxvaerd1990}.

We employed a cubic bounding box with periodic boundary conditions containing $N=2000$ particles (200 chains).  The simulations were performed using the MD code RUMD~\cite{RUMD}, which is optimized for GPU computing~\cite{Bailey2015}. In some of the figures below we compare our results to simulations of the LJC with rigid bonds. Most of the latter data come from paper I, in which the details of the simulation of the rigid-bond chains can be found~\cite{Veldhorst2014}.

\subsection{Background of the model}
The Lennard-Jones chain was first simulated by Kremer and Grest~\cite{Grest1986, Kremer1988, Kremer1990}, who used it as a coarse-grained model to study the properties of polymeric liquids. The particles in the chain correspond to groups of atoms, like one or several $\mathrm{CH}_n$ units in an alkane, or one or several monomers in a polymer. For this reason the Lennard-Jones particles in the chain are referred to as ``segments''.

Starting at the end of the 90's, extensive simulations of the model have been done to investigate the behavior of polymer melts around the glass transition~\cite{Bennemann1998, Bennemann1999, Bennemann1999b, Bennemann1999c, Bennemann1999d, Binder1999}. At that time the model had already undergone some changes compared to the original version. The main difference was that the new simulations did not cut and shift the potential at the minimum, but also included the attractive part of the Lennard-Jones potential~\cite{Kopf1997}, whereas the earlier versions cut and shifted the potential at the minimum. A second difference was that Kremer and Grest~\cite{Grest1986} did not use Molecular Dynamics, but Langevin Dynamics, which includes a stochastic force similar to what is done when simulating an implicit solvent~\cite{Schlick}.

In the original version of the LJC model~\cite{Grest1986}, as well as in many later simulations where the attractive part of the LJ potential is taken into account~\cite{Bennemann1998, Bennemann1999, Bennemann1999b, Bennemann1999c, Bennemann1999d, Binder1999}, bonds were modeled with the following finitely extensible nonlinear elastic (FENE) potential
\begin{equation}
  \upsilon(r) = -0.5kr_{max}^2 \ln\left[1-\left(\dfrac{r}{r_{max}}\right)^2\right]\,.
\end{equation}
Here, $r_{max}$ is the maximum length of the bond at which the potential diverges. Since the FENE potential is purely attractive, it is used in addition to the Lennard-Jones potential. This is in contrast to the harmonic-bond chains used in this work, where there is no Lennard-Jones interaction between bonded particles. The combination of the FENE and the Lennard-Jones potentials results in a potential minimum of approximately $0.96\sigma$.

Some more recent simulations of Lennard-Jones chains have used harmonic bonds~\cite{Goel2008, Galliero2009a, Riggleman2009, Riggleman2010, Galliero2011, Shavit2013} and rigid bonds~\cite{Galliero2009b, Galliero2011, Veldhorst2014}. In these studies, the bond length is always set to $\sigma$ for both the harmonic and the rigid bonds. In the case of harmonic springs, the spring constant is always $k=3000\varepsilon/\sigma^2$, which is rather stiff and leads to narrow bond length distributions. Therefore the stiff harmonic bonds and rigid bonds are often considered to be equivalent, at least concerning the phase diagram of the LJC phase diagram~\cite{Johnson1994, Galliero2011}. The phase diagram of the LJC model with FENE bonds is not expected to be the same due to the different bond lengths. Moreover, the shorter bond lengths will have a significant effect on the molecular structure. We therefore decided to investigate the effect of non-rigid bonds by comparing our previous results for rigid bonds~\cite{Veldhorst2014} with new simulations of the LJC model with harmonic bonds.

\subsection{Effects of bond type}
\begin{figure}
  \centering
  \includegraphics[width=\figwidth]{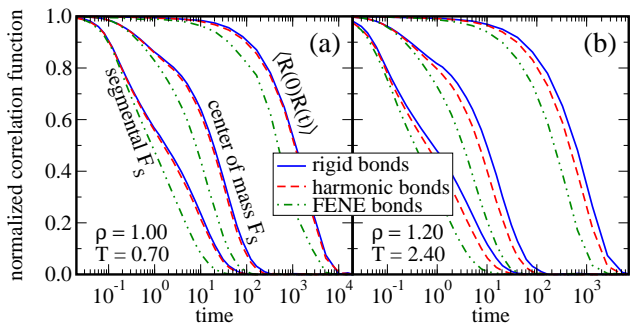}
  \caption{Comparison of the dynamics of the Lennard-Jones chain model with rigid bonds (blue lines) and harmonic bonds (red dashed lines). The self-intermediate scattering function $F_S(q,t)$ with $q=7.09$ of the segments and the center of mass, and the autocorrelation function of the end-to-end vector $\left\langle\vect{R}(0)\vect{R}(t)\right\rangle$ are plotted. (a) At a standard density ($\rho=1.0$) the harmonic and rigid bonds give practically the same dynamics, while the LJC model with FENE bonds (green dotted lines) has faster dynamics due to the smaller bond length. (b) At higher density and temperature, the chains with harmonic bonds have slightly faster dynamics than the chains with rigid bonds. The two state points in (a) and (b) are isomorphic to one another for the rigid-bond chains~\cite{Veldhorst2014}.}\label{fig:dynamics-bonds}
\end{figure}

We compare the dynamics of the LJC models with different bond types in Fig.~\ref{fig:dynamics-bonds}, where we plot the intermediate scattering function of the segments and the center of mass, as well as the autocorrelation function of the end-to-end vector $\left\langle\vect{R}(0)\vect{R}(t)\right\rangle$. Figure~\ref{fig:dynamics-bonds}(a) shows the dynamics at a standard dense liquid state point. As mentioned earlier~\cite{Johnson1994, Galliero2011}, the harmonic and rigid bonds give indistinguishable dynamics at this state point. Nevertheless, at another state point with higher density and temperature, the dynamics of the chains with harmonic and rigid bonds start to differ, as shown in Fig.~\ref{fig:dynamics-bonds}(b), so the two models with different bond types cannot be considered to be equivalent for all state points.

The two state points investigated in Fig.~\ref{fig:dynamics-bonds}(a) and~(b) have been shown in paper I to be isomorphic to each other for the LJC liquid with rigid bonds, i.e., they have to a very good approximation the same dynamics and (intermolecular) structure. We also showed there that the isomorph in the phase diagram was well described by the condition
\begin{equation}\label{eq:hrho-constant}
  \dfrac{h(\rho)}{T} = \mathrm{Const.}\,,
\end{equation}
with $h(\rho) = 2\rho^{5.06}-\rho^{2.61}$ determined by empirical scaling~\cite{Veldhorst2014}. The fact that the LJC model with harmonic bonds does not have the same dynamics as the model with rigid bonds indicates one of two things: \emph{either} the model with harmonic bonds conforms to the isomorph theory but with isomorphs described by a different $h(\rho)$, \emph{or} the model with harmonic bonds does not conform to the isomorph theory. Earlier it has been shown that the LJC liquid with harmonic bonds obeys power-law density scaling~\cite{Galliero2011}, which is a good approximation to isomorphic scaling in small density ranges. Moreover, the model has also been shown to obey Rosenfeld's excess-entropy scaling~\cite{Goel2008, Voyiatzis2013}, which is in agreement with the isomorph theory. These facts indicate that the LJC liquid with harmonic bonds could be described by the isomorph theory, albeit with an $h(\rho)$ that is slightly different from the rigid bond chains. As we shall see, this is not the case, although the model does have curves of invariant dynamics and structure.

\begin{figure}
  \centering
  \includegraphics[width=\figwidth]{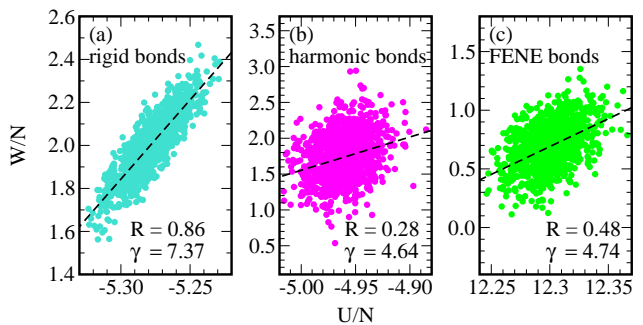}
  \caption{Scatter plots of the potential energy $U$ and virial $W$ equilibrium fluctuations for the LJC model with different bond types. Where the rigid bond simulations (a) have quite strong $UW$ correlations, the use of harmonic bonds (b) destroys the correlations. The FENE bonds (c) have a slightly higher correlation coefficient than the harmonic bonds. For all three bond types, the data were obtained at the state point $\rho = 1.0$, $T=0.7$.}\label{fig:fluctuations-bonds}
\end{figure}

So far it seems like the bond type only has a small effect on the behavior of the liquid. This is, however, not the case when looking at a prominent property of R liquids. Figure~\ref{fig:fluctuations-bonds} shows scatter plots of the fluctuations of the potential energy $U$ and the virial $W$ for the three different bond types. It is immediately apparent that the bond type has a big effect on the correlations of the these two quantities. For the rigid bonds it was already shown in paper I that the correlation coefficient of the LJC liquid is $R=0.86$ at the state point  $\rho=1.0$, $T= 0.7$. Fig.~\ref{fig:fluctuations-bonds}(b) shows that if the same state point is simulated with harmonic bonds, the correlations disappear almost completely ($R=0.28$), even though the dynamics do not change (see Fig.~\ref{fig:dynamics-bonds}(a)). We also show data for the FENE bonds in Fig.~\ref{fig:fluctuations-bonds}(c). Here the correlations are stronger ($R=0.48$) than in the system with harmonic bonds, but still far from the value of the chains with rigid bonds.

Both the strong $UW$ correlations and the existence of isomorphs in a liquid's phase diagram are properties of an R liquid, and it has been shown that $UW$ correlations imply the existence of isomorphs and vice versa~\cite{paper4}. In view of this it is puzzling that earlier investigations of the model with harmonic bonds have shown that it obeys power-law density scaling~\cite{Galliero2011} and Rosenfeld's excess-entropy scaling~\cite{Goel2008, Voyiatzis2013}, indicating that it \emph{is} a simple liquid, while the strong $UW$ correlations are absent, indicating that it \emph{is not} an R liquid.

\section{Dynamics and structure along a configurational adiabat\label{sec:configurational_adiabat}}
Isomorphs are often created by keeping excess entropy constant (Sec.~\ref{sec:theory}), and we create a configurational adiabat using the same method. We circumvent the time-consuming calculation of the excess entropy by using Eq.~(\ref{eq:gamma-rhoT}) to keep $S_{ex}$ constant (without knowing its value). We do this by performing an initial equilibrium simulation at the state point $(\rho_1,T_1) = (1.0,0.7)$. $\gamma$ is then calculated from the fluctuations in $U$ and $W$ using Eq.~(\ref{eq:gamma-fluctuations}). Note that this is possible even if the liquid does not obey the isomorph theory, and the fluctuations in $U$ and $W$ are not correlated~\cite{paper1, paper4}. We choose a density $\rho_2$ for the new state point, which is close to the density of the initial state point. It is then possible to calculate the temperature $T_2$ at this new density for which the excess entropy is identical to the first state point, by rewriting Eq.~(\ref{eq:gamma-rhoT}) to $T_2 = T_1(\rho_2/\rho_1)^\gamma$. The procedure is repeated several times by doing an equilibrium simulation at the new state point to calculate a new value of $\gamma$ and find a new state point on the adiabat. It is important to choose the change in density $|\rho_2-\rho_1|$ small enough because $\gamma$ may change with density. This can be verified by making sure that a further decrease of this density difference does not have a significant effect on the result. Here, the change in density was 0.02, and we obtained a set of state points with densities ranging from 0.96 to 1.08.

\begin{figure}
  \centering
  \includegraphics[width=\figwidth]{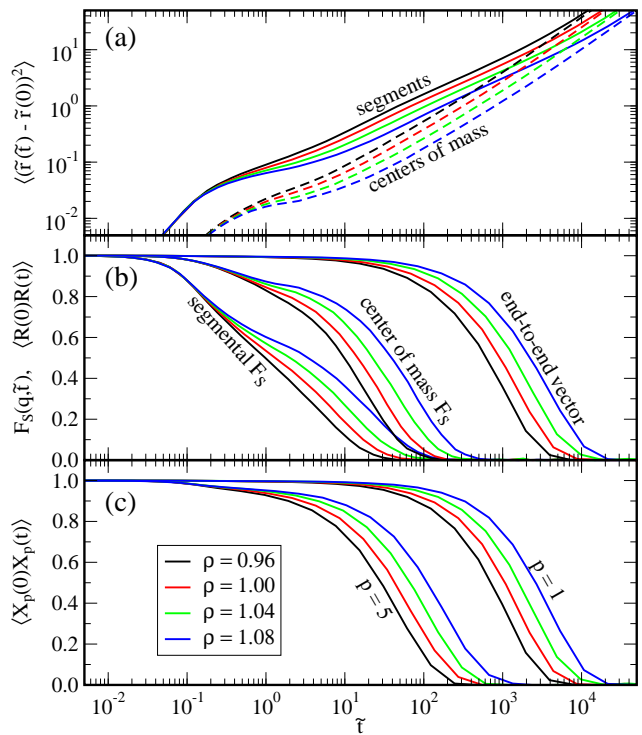}
  \caption{Dynamics at four state points with the same excess entropy, as a function of reduced time. (a) The mean square displacement of the segments and the center of mass of the chain. (b) The segmental and center-of-mass incoherent intermediate scattering function $F_S(q, \tilde{t})$ and the autocorrelation of the end-to-end vector $\left\langle\vect{R}(0)\vect{R}(t)\right\rangle$. The length of the scattering vector was kept constant in reduced units as $q=\tilde{q}\rho^{1/3}$ with $\tilde{q}=7.09$ approximately at the main peak of the static structure factor. (c) The autocorrelation functions of the first and the fifth Rouse modes. All these dynamical quantities differ at the four state points, showing that for the LJC model with harmonic bonds, the dynamics are not a function of the excess entropy alone.}\label{fig:ljch_im4_dynamics}
\end{figure}

For R~liquids, a configurational adiabat is an isomorph, and therefore the dynamics are invariant along a configurational adiabat when plotted in reduced units. We test this in Fig.~\ref{fig:ljch_im4_dynamics} where we plot various dynamical quantities of the harmonic bond LJC model for state points on the configurational adiabat. The figure contains (a) mean square displacements, (b) incoherent intermediate scattering functions $F_S(q,t)$ and autocorrelation functions of the end-to-end vector $\left\langle\vect{R}(0)\vect{R}(t)\right\rangle$, and (c) Rouse-mode autocorrelation functions $\left\langle\vect{X}_p(0)\vect{X}_p(t)\right\rangle$ for $p=1,5$. All these quantities, which probe the segmental dynamics of the individual LJ particles, as well as the chain dynamics, are clearly changing along the configurational adiabat. 

\begin{figure}
  \centering
  \includegraphics[width=\figwidth]{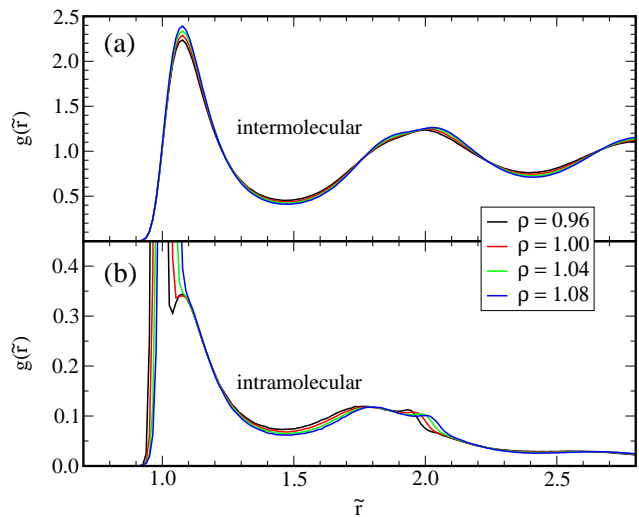}
  \caption{The radial distribution function $g(\tilde{r})$ along the configurational adiabat, in reduced units. We split the total (segmental) $g(\tilde{r})$ into (a) intermolecular contributions and (b) intramolecular contributions. For the intermolecular $g(\tilde{r})$ there is a reasonable collapse, but for the intramolecular $g(\tilde{r})$ there are clear differences, especially around the bond length.}\label{fig:ljch_im4_rdf}
\end{figure}

The isomorph theory predicts that the structure of the liquid is invariant along a configurational adiabat. For completeness we also test this with the radial distribution functions plotted in Fig.~\ref{fig:ljch_im4_rdf}. The radial distribution functions are split in the intermolecular contributions of particle pairs in different molecules and intramolecular contributions from pairs in the same molecule. The reason is that we have previously shown that only the intermolecular contribution is invariant on the isomorph for the LJC model with rigid bonds~\cite{Veldhorst2014}. For the chains with harmonic bonds we find qualitatively the same results on the configurational adiabat; the average bond length does not scale with density, and is therefore not invariant in reduced units. The intramolecular structure is thus not the same at different densities. The intermolecular structure is reasonably invariant, since the first peak is exactly at the same position for the tested densities. However, it should be noted the height of the first peak of $g(r)$ changes more than what was found for the LJC with rigid bonds on the isomorph~\cite{Veldhorst2014}.

From the results presented so far we conclude that the LJC model with harmonic bonds does not obey the isomorph theory or Rosenfeld's excess-entropy scaling, because the dynamics are not invariant on a curve of constant excess entropy. This is in disagreement with previous results, that have shown that the LJC with flexible bonds does obey Rosenfeld's excess-entropy scaling~\cite{Goel2008, Voyiatzis2013}. The reason for this discrepancy is the way that the excess entropy is calculated. The excess entropy can been approximated in different ways, the most exact of which is thermodynamic integration. The simplest approximation is the pair entropy $S_2$ which can easily be calculated from the radial distribution function, but in this case the bonded particle pairs are excluded in the calculation of the $g(r)$. A third method that is commonly used employs an equation of state developed using Self-Associating Fluid Theory~\cite{Johnson1994}. Voyiatzis \textit{et al.}~\cite{Voyiatzis2013} have compared the effect of the different entropy approximations on the applicability of Rosenfeld's excess-entropy scaling of Lennard-Jones chains with harmonic bonds. They found that the scaling works best when $S_2$ is used, ignoring the bonded particle pairs. If the bonded particle pairs were not removed in the calculation of the entropy, as is the case in the thermodynamic integration, the transport coefficients could not be collapsed on a single curve. Our method of identifying isomorphs using Eq.~(\ref{eq:gamma-fluctuations}) and (\ref{eq:gamma-rhoT}) avoids these problems.

\section{Identification of a pseudoisomorph\label{sec:pseudoisomorph}}
Galliero \textit{et al.}~\cite{Galliero2011} have  shown that reduced viscosities of the LJC model with harmonic bonds can be scaled approximately onto a single curve that is a function of $h(\rho)=\rho^\gamma/T$. This may seem surprising given our previously mentioned result that power-law density scaling does \emph{not} work for the LJC model, but in Ref.~\cite{Galliero2011} the collapse is not perfect, and the extent of the densities investigated is not mentioned. Nevertheless, the results of Galliero \textit{et al.} indicate that a curve exists which is similar to an isomorph, along which dynamics and structure are invariant.

For an R liquid, the curve in the phase diagram described by $h(\rho)$ is called an isomorph. Moreover, not only the reduced viscosity, but all dynamical measures, the structure in reduced units, and the excess entropy are constant on this curve. In this section we construct a curve of invariant dynamics and test to what degree it has the properties of an isomorph. Since we have shown in the previous section that it cannot be a proper isomorph because the excess entropy is not constant, we call this curve of invariant dynamics a \emph{pseudo}isomorph. We construct the pseudoisomorph by empirical density scaling of the segmental relaxation times, but unlike Ref.~\cite{Galliero2011} do not make any assumption about the functional form of the scaling function $h(\rho)$.

\begin{figure}
  \centering
  \includegraphics[width=\figwidth]{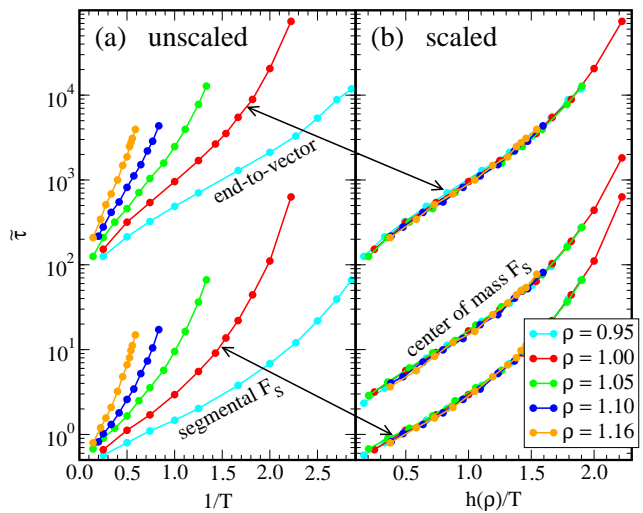}
  \caption{Empirical density scaling of the (reduced) relaxation time for the harmonic bond chains. On the left the unscaled data from the segmental incoherent intermediate scattering function $F_S(q,\tilde{t})$ ($q=7.09\rho^(1/3)$) and the autocorrelation of the end-to-end vector $\left\langle\vect{R}(0)\vect{R}(t)\right\rangle$ are shown for different isochores. On the right, the segmental relaxation data have been scaled by hand to obtain $h(\rho)$, and two other measures of relaxation times have been scaled by the same factor.}\label{fig:ljch_scaling}
\end{figure}

Relaxation times were determined from the incoherent intermediate scattering function of the segments and the center of mass, and from the correlation function of the end-to-end vector. We defined the relaxation time as the time after which the normalized correlation function has decayed to 0.2. Unscaled reduced relaxation times of the segmental $F_S$ and the end-to-end vector are shown in Fig.~\ref{fig:ljch_scaling}(a) for five isochores with densities ranging from $\rho=0.95$ to $\rho=1.16$. It was not possible to go to higher relaxation times (lower temperatures) due to crystallization at higher densities and phase separation or negative pressures at lower densities. Nevertheless a fairly large range of temperatures could be reached for some densities.

 With standard power-law density scaling, a function $h(\rho)= \rho^{\gamma_s}/T$ is found which collapses the isochoric data when plotted versus $\rho^{\gamma_s}$, where $\gamma_s$ is a material-specific constant. Instead, we scale each isochore to collapse the segmental relaxation times onto the $\rho=1.00$ isochore. Thus for each isochore, a scalar $h$ was chosen by hand to collapse the segmental relaxation times as functions of $h/T$. The value of the scaling parameter $h$ was found independently for each isochore from the segmental relaxation times. The results of the scaling in Fig.~\ref{fig:ljch_scaling}(b) show a good collapse for all three measures of the relaxation time, even though only the segmental relaxation times were used in the scaling procedure. It was found earlier with power-law density scaling that both the segmental and chain dynamics follow the same scaling~\cite{Roland2004b, Casalini2005}, and this has also been confirmed for an all-atom polymer model~\cite{Tsolou2006}.

\begin{figure}
  \centering
  \includegraphics[width=\figwidth]{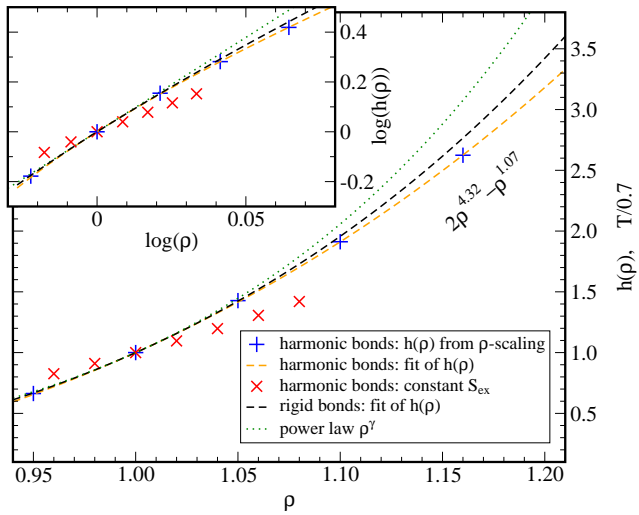}
  \caption{The values of the harmonic bond $h(\rho)$ obtained by empirical scaling (blue crosses) compared to the rigid bond $h(\rho)$ (black dashed line). The data for the harmonic bond chains have been fitted to obtain $h(\rho) = 2\rho^{4.32}-\rho^{1.07}$ (orange dashed line). We included the shape of the configurational adiabat (red crosses) by plotting the temperature at each density divided by the temperature at $\rho=1$. The green dots show a curve of constant $\gamma=7.57$. The inset shows the same but in a log-log plot, where a constant $\gamma$ leads to a straight line.}\label{fig:ljch_TvsRho}
\end{figure}

The values for $h(\rho)$ that were obtained from the scaling are plotted in Fig.~\ref{fig:ljch_TvsRho} (blue crosses). The data for the harmonic springs are compared with the fit of $h(\rho)$ for the rigid bond chains (black dashed line) from paper I~\cite{Veldhorst2014}. There is a small but significant difference in the shapes of $h(\rho)$ for the two models. The difference is most obvious at high density, where the harmonic bond $h(\rho)$ is lower than the rigid bond $h(\rho)$. This is in agreement with the data in Fig.~\ref{fig:dynamics-bonds}(b), which show that the dynamics of the chains with harmonic bonds is faster at high density. To keep the dynamics invariant on a pseudoisomorph, the temperature on the isomorph of the harmonic bond chain should be lower at high densities than for the isomorph of the LJC with rigid bonds. Recall that on an isomorph, $T\propto h(\rho)$ (see Eq.~(\ref{eq:hrho-constant}), so also $h(\rho)$ should be lower for the harmonic bond chains. As in the previous paper, we have fitted the data to a function of the form $h(\rho)=2\rho^\alpha-\rho^\beta$. The resulting function $h(\rho)=2\rho^{4.32}-\rho^{1.07}$ is shown as the dashed orange line.

The inset of Fig.~\ref{fig:ljch_TvsRho} shows the same data in a log-log plot. A power-law $h(\rho)=\rho^\gamma$ corresponds to a straight line in this plot (green dots). Our data show that this is not a good description of the data, which means that power-law density scaling is an approximation that only works for the smaller density changes (5\%), confirming previous findings of ours~\cite{Bohling2012}.

\begin{figure}
  \centering
  \includegraphics[width=\figwidth]{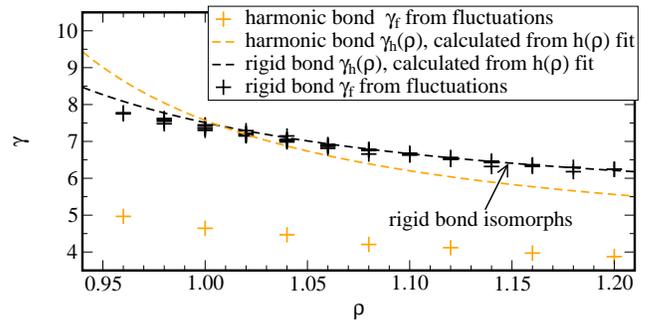}
  \caption{$\gamma_h(\rho)$ for the harmonic bond chains (orange dashed line) calculated from the $h(\rho)$ fit in Fig.~\ref{fig:ljch_TvsRho}. This is compared to the rigid bond $\gamma_h(\rho)$ (black dashed line) calculated from $h(\rho)$ and the rigid bond $\gamma$ values calculated from the $UW$ fluctuations (Eq.~(\ref{eq:gamma-fluctuations})). For harmonic bonds, the $\gamma_f$ values calculated from the fluctuations (orange crosses) are significantly lower. These have been calculated at state points on the configurational adiabat.}\label{fig:ljch_gamma-rho}
\end{figure}

For liquids that obey the isomorph theory, $h(\rho)$ describes the shape of configurational adiabats. From Eq.~(\ref{eq:gamma-rhoT}) it then follows that $\gamma$ as calculated from the fluctuations (from now on denoted by $\gamma_f$) is the same as the logarithmic derivative of $h(\rho)$~\cite{Ingebrigtsen2012}. For a pseudoisomorph this may not hold, since $h(\rho)$ does not describe a configurational adiabat. It is however still possible to calculate the logarithmic derivative of $h(\rho)$ for the pseudoisomorph as
\begin{equation}\label{eq:gamma-rho}
  \gamma_h(\rho) = \dfrac{\ud\ln h(\rho)}{\ud \ln \rho} \neq \gamma_f\,.
\end{equation}
The result of this is plotted in Fig.~\ref{fig:ljch_gamma-rho} (orange dashed line). Our values are consistent with the value $\gamma \approx 6.5$ that Galliero \textit{et al.} found using power-law density scaling. Comparing $\gamma_h(\rho)$ of the harmonic bonds with $\gamma_h(\rho)$ for rigid bonds (dashed black line) we see that $\gamma$ has a similar magnitude, but a stronger density dependence for the chains with harmonic bonds. For rigid bonds $\gamma_h(\rho)$ and $\gamma_f$ are identical. In contrast, the chains with harmonic bonds, using Eq.~(\ref{eq:gamma-fluctuations}) to calculate $\gamma_f$ from the fluctuations (orange crosses), gives $\gamma_f$ values that are much lower than the $\gamma(\rho)$ from the fitted $h(\rho)$. This confirms that the pseudoisomorph is \emph{not} a configurational adiabat.

After having established that the pseudoisomorph is not a configurational adiabat, we test to which degree the former has other isomorph invariants. We obtain a set of pseudoisomorphic state points from the fitted expression of $h(\rho)$ using $T = T_0(2\rho^{4.32}-\rho^{1.07})$ (Eq.~(\ref{eq:gamma-rho})), with $T_0=0.7$ the temperature at $\rho=1$. We use densities from 0.96 to 1.20, creating a pseudoisomorph along which the density changes by 25\%.

\begin{figure}
  \centering
  \includegraphics[width=\figwidth]{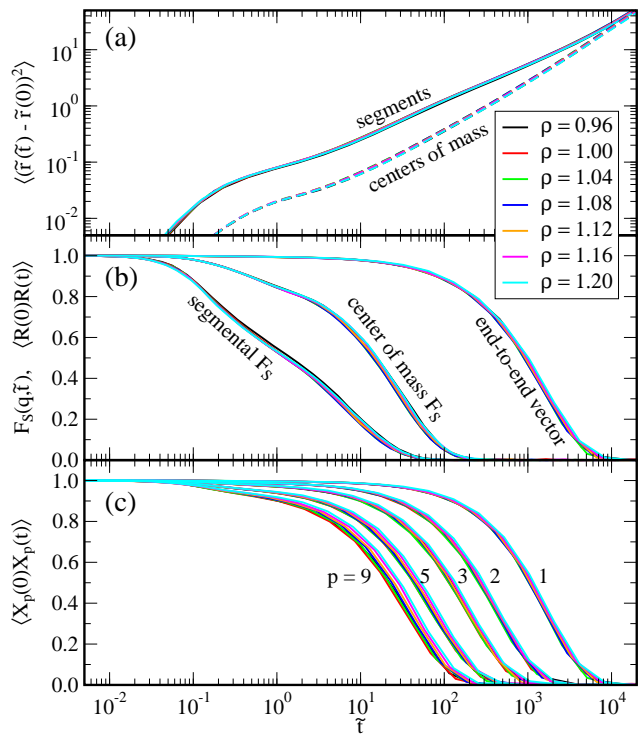}
  \caption{The dynamics at the pseudoisomorphic state points. The pseudoisomorphic state points were chosen to have the same relaxation time of the segmental incoherent intermediate scattering function (procedure described in text). (a) The mean square displacement of the segments (solid lines) and the centers of mass (dashed lines) (b) The incoherent intermediate scattering function $F_s(q,t)$ of the segments and the centers of mass, and the autocorrelation function of the end-to-end vector $\left\langle\vect{R}(0)\vect{R}(t)\right\rangle$. (c) Some of the autocorrelation functions of the Rouse modes $\left\langle\vect{X}_p(0)\vect{X}_p(t)\right\rangle$ in reduced units for pseudoisomorphic state points. The collapse is good for the lower modes, but for the higher modes some deviation is seen. The dynamics are invariant on the pseudoisomoph, while they ar not at the configurational adiabat (Fig.~\ref{fig:ljch_im4_dynamics}).}\label{fig:im8_dynamics}
\end{figure}

In Fig.~\ref{fig:im8_dynamics} we plot different dynamical quantities in reduced units at the pseudoisomorphic state points. These include the segmental and center-of-mass mean square displacements and incoherent intermediate scattering functions, and chain specific quantities like the orientational autocorrelation of the end-to-end vector and Rouse-mode autocorrelation functions. By definition the relaxation times of the segmental intermediate scattering function are invariant on the pseudoisomorph. The data show that the entire shape of the relaxation functions is the same for all quantities. There is only a slight deviation at the high Rouse modes, which correspond to movements in short subchains. This is identical to what was found in paper I for chains with rigid bonds~\cite{Veldhorst2014}. The lower Rouse modes correspond to larger (sub)chains, and these are as invariant as the other quantities in the figure. All correlation functions are found to be much more invariant on the pseudoisomorph than on the configurational adiabat (Fig.~\ref{fig:ljch_im4_dynamics}).

\begin{figure}
  \centering
  \includegraphics[width=\figwidth]{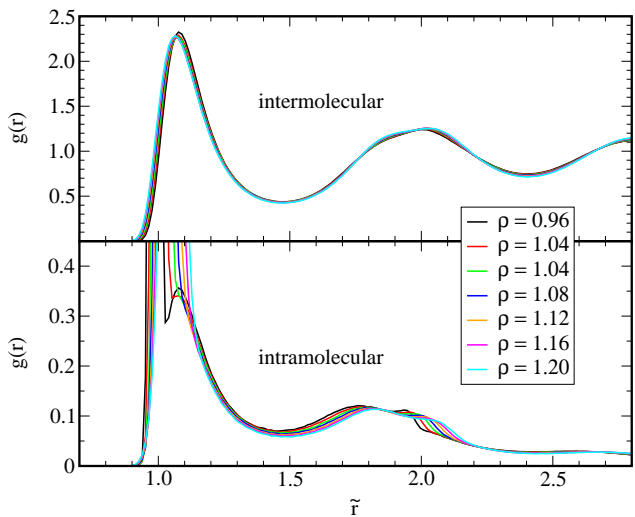}
  \caption{Radial distribution functions $g(r)$ in reduced units on the pseudoisomorph. (a) The intermolecular $g(r)$ is invariant on the pseudoisomorph. (b) The intramolecular $g(r)$ is not invariant. This is mainly due to the bonded particle pairs, but there are also significant differences at larger distances. The structure is more invariant on the pseudoisomorph than on the configurational adiabat (Fig.~\ref{fig:ljch_im4_rdf})}.\label{fig:ljch_im8_rdf}
\end{figure}

Next we investigate whether the structure of the liquid is also invariant on the pseudoisomorph. We plot the radial distribution functions of the pseudoisomorphic state points in Fig.~\ref{fig:ljch_im8_rdf}. As in Fig.~\ref{fig:ljch_im4_rdf} the radial distribution function is split into an intermolecular contribution (a) and an intramolecular contribution (b). As for the rigid-bond isomorphs (paper I), we find that the intermolecular structure is invariant on the pseudoisomorph, while the intramolecular structure is not. The main reason for this is the bonded particle pairs in the molecule.

\begin{figure}
  \centering
  \includegraphics[width=\figwidth]{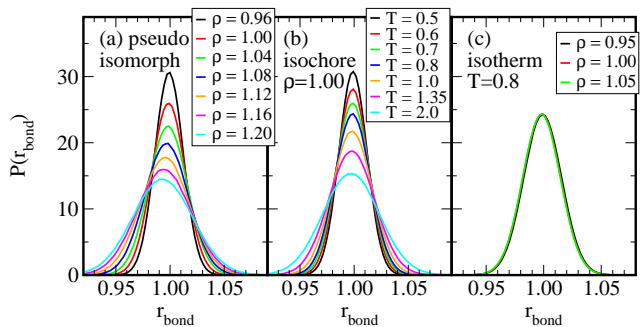}
  \caption{Bond-length distributions along the pseudoisomorph, an isochore, and an isotherm. The width of the bond-length distribution changes along the pseudoisomorph and isochore, but is constant on an isotherm (equipartition). The average bond length changes slightly on the pseudoisomorph. Note the that in this figure, the bond length is \emph{not} given in reduced units.}\label{fig:ljch_im8_bonds}
\end{figure}

From Fig.~\ref{fig:ljch_im8_rdf}(b) it is clear that the behavior of the bonds is complicated and varies on the pseudoisomorph, so we now investigate this further. For the LJC model with rigid bonds, the bonds cannot follow the scaling, because their lengths are kept constant in normal units, meaning that they change in reduced units. Nevertheless, their behavior on the isomorph was rather trivial; they show up as delta functions in the $g(r)$~\cite{Veldhorst2014}. Fig.~\ref{fig:ljch_im8_bonds}(a) shows that for the harmonic bonds, also the width of the distribution changes on the pseudoisomorph. The bond length distributions on the isochore and isotherm (Fig.~\ref{fig:ljch_im8_bonds}(b) and (c)) show that the width only depends on temperature, as expected from the equipartition theorem. The average bond length changes slightly on the pseudoisomorph, but not enough to be invariant in reduced units. The fact that at high temperatures the chains with harmonic bonds have faster dynamics than the chains with rigid bonds (see Fig.~\ref{fig:dynamics-bonds}(b)) may be related to the wider bond length distribution, making it easier for the segments and the chain to cross barriers.

\begin{figure}
  \centering
  \includegraphics[width=\figwidth]{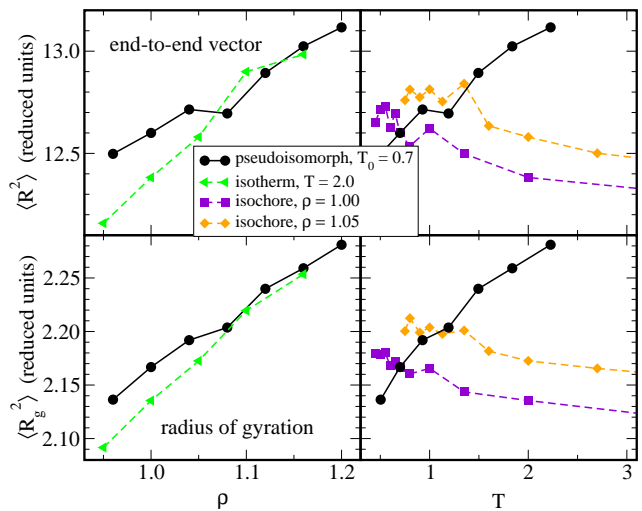}
  \caption{The mean square end-to-end vector $\langle\vect{R}^2\rangle$ and mean square radius of gyration $\langle R_g^2\rangle$ on the pseudoisomorph, compared to the isotherm and isochore. These measures of the molecule size seem only dependent on density, and are not invariant on the pseudoisomorph.}\label{fig:ljch_im8_R0R0}
\end{figure}

We plot two measures of the molecular size in Fig.~\ref{fig:ljch_im8_R0R0}; the mean square end-to-end vector $\langle\vect{R}^2\rangle$ and the mean square radius of gyration $\langle R_g^2\rangle$, both in reduced units. The size of the chains is not invariant on the pseudoisomorph, and the molecular sizes seem to depend only on density, since it is almost constant on the isochores. The data are very similar to those found for rigid bonds~\cite{Veldhorst2014}. It seems that the effect of temperature is slightly larger for the harmonic bonds, which we attribute to the temperature dependence of the the bond length distributions.

\section{Discussion and outlook\label{sec:conclusion}}
Our analysis of the structure and dynamics shows that the LJC model with harmonic bonds has pseudoisomorphs, which are very similar to the isomorphs of the LJC model with rigid bonds. Dynamics and structure are invariant on these curves, except when very local intramolecular contributions are considered, such as the higher Rouse modes and the separation between nearest and next-nearest neighbors. The pseudoisomorphs of the chains with harmonic bonds have a different shape from the isomorphs of the chains with rigid bonds, especially at high densities and temperatures. This difference is presumably caused by the wider bond length distributions for flexible bonds at high temperatures. 

The main differences between the pseudoisomorphs in this paper and the isomorphs in paper~I are related to the fluctuations in the energy and the pressure. The $UW$ correlations are weak, and $\gamma$ as calculated from the $UW$ fluctuations does not agree with the logarithmic slope of the pseudoisomorph in the $\rho,T$ phase diagram. Therefore the excess entropy is not constant on the pseudoisomorph. The LJC liquid with harmonic bonds does thus not obey the isomorph theory, even though the same model with rigid bonds does and has similar dynamics and structure.

This also means that the liquid with harmonic bonds does not obey Rosenfeld's excess-entropy scaling when the bonds are flexible, i.e., the excess entropy does not control the relaxation time. This is in disagreement earlier results, where the chains with harmonic bonds have been shown to obey Rosenfeld's excess-entropy scaling~\cite{Goel2008, Voyiatzis2013}. The disagreement of these previous results with our conclusion may be ascribed to the fact that the collapse of the data was not exact in these previous studies. Moreover, Voyiatzis \emph{et~al.} have shown that excess entropy scaling works best when the entropy is approximated by the pair entropy $S_2$ and the bonded particle pairs are ignored~\cite{Voyiatzis2013}. We conclude that the flexible bonds contribute to the entropy of the system, but this contribution is not related to the (long-time) dynamics of the system.

The pseudoisomorphs in this article have been identified using empirical scaling. This is inferior to real isomorphs which can be constructed using Eqs.~(\ref{eq:gamma-fluctuations}) and (\ref{eq:gamma-rhoT}). It would be desirable to be able to find the pseudoisomorphs without reverting to empirical scaling, and work is in progress with this aim.

\acknowledgments{
The centre for viscous liquid dynamics ``Glass and Time'' is sponsored by the Danish National Research Foundation via Grant No. DNRF61.
}


\end{document}